\documentclass[%
 reprint,
 amsmath,amssymb,
 aps,
]{revtex4-1}

\usepackage{graphicx}% Include figure files
\usepackage{dcolumn}% Align table columns on decimal point
\usepackage{bm}% bold math
\begin{document}

%\preprint{APS/123-QED}

\title{Size Scaling of Velocity Field in Granular Flows through Apertures}% Force line breaks with \\
%\thanks{A footnote to the article title}%
\author{Gaoke Hu$^{1,4}$}{}
\author{Ping Lin$^2$}{}
\author{Yongwen Zhang$^{1,4}$}
\author{Liangsheng Li$^{3}$}{}
\author{Lei Yang$^2$}
\author{Xiaosong Chen$^{1,4}$}%
\email{Email address: chenxs@itp.ac.cn }
\address{$^1$Institute of Theoretical Physics, Key Laboratory of Theoretical Physics, Chinese Academy of Science, P.O. Box 2735, Beijing, 100190, China\\
$^2$ Institute of Modern Physics, Chinese Academy of Sciences, 509 Nanchang Road, Lanzhou 730000, China\\
$^3$ Science and Technology on Electromagnetic Scattering Laboratory, Beijing 100854, China\\
$^4$School of Physical Sciences, University of Chinese Academy of Science,  No. 19A Yuquan Road, Beijing 100049, China
}

%\collaboration{CLEO Collaboration}%\noaffiliation

\date{\today}% It is always \today, today,
             %  but any date may be explicitly specified
%\date{August 10, 2010}%
\begin{abstract}
For vertical velocity field $v_{\rm z} (r,z;R)$ of granular flow through an aperture of radius $R$, we propose a size scaling form $v_{\rm z}(r,z;R)=v_{\rm z} (0,0;R)f (r/R_{\rm r}, z/R_{\rm z})$ in the region above the aperture. The length scales $R_{\rm r}=R- 0.5 d$ and $R_{\rm z}=R+k_2 d$, where $k_2$ is a parameter to be determined and $d$ is the diameter of granule. The effective acceleration, which is derived from $v_{\rm z}$, follows also a size scaling form $a_{\rm eff} = v_{\rm z}^2(0,0;R)R_{\rm z}^{-1} \theta (r/R_{\rm r}, z/R_{\rm z})$. For granular flow under gravity $g$, there is a boundary condition $a_{\rm eff} (0,0;R)=-g$ which gives rise to $v_{\rm z} (0,0;R)= \sqrt{ \lambda g R_{\rm z}}$ with $\lambda=-1/\theta (0,0)$. Using the size scaling form of vertical velocity field and its boundary condition, we can obtain the flow rate $W =C_2  \rho \sqrt{g } R_{\rm r}^{D-1} R_{\rm z}^{1/2} $, which agrees with the Beverloo law when $R \gg d$.  The vertical velocity fields $v_z (r,z;R)$ in three-dimensional (3D) and two-dimensional (2D) hoppers have been simulated using the discrete element method (DEM) and GPU program. Simulation data confirm the size scaling form of $v_{\rm z} (r,z;R)$ and the $R$-dependence of $v_{\rm z} (0,0;R)$.  
\begin{description}
%\item[Usage]
\item[PACS numbers]
45.70.Mg
%May be entered using the \verb+\pacs{#1}+ command.
%\item[Structure]

\end{description}
\end{abstract}

%\pacs{Valid PACS appear here}% PACS, the Physics and Astronomy
                             % Classification Scheme.
%\keywords{Suggested keywords}%Use showkeys class option if keyword
                              %display desired
\maketitle

%\tableofcontents

%Granular matter wideliny exists on the earth and has been widespread used in our daily lives.
Granular materials flowing through apertures show many unusual physical  properties and have been studied extensively for decades\citep{Nedderman1982}.  Depending on the size of apertures, granular flows can have three patterns which are continuous, intermittent, and jammed respectively. Contrary to fluids, flow rate of granular materials in the continuous pattern does not depend on the height of the granular layer above the aperture. Beverloo \emph{et al.} \citep{Beveloo} proposed an empirical expression of granular material flow rate  $W$ driven by gravity as
\begin{eqnarray}
W =C\rho \sqrt{g}(R-kd)^{D- 1/2},
\end{eqnarray}
where $g$ is the gravitational acceleration, $\rho$ is  the bulk density, $R$ is the radius of aperture, $d$ is the granule diameter, and $D$ is the dimensionality of hopper. $C$ and $k$ are two fitted parameters. The Beverloo law was related to a hypothesis of free-fall arch, which was introduced by Hagen \cite{Tighe2007} and developed lately by Brown and Richards \cite{Brown1961}. Velocities of the granules above the arch are considered to be negligible. Below the arch, granules fall freely \cite{Tighe2007,Brown1961,Nedderman1982,Janda,Mankoc2007}. However,  the acceleration profiles obtained experimentally by Rubio-Largo  \emph{et al.} are against the existence of free-fall arch described by the Heavisde function \cite{Rubio}. 

For granular flow on conveyor belt, Bao \emph{et al.} \cite{Bao2003} found that 2D flow rate $W$ is proportional to $R$ when the velocity of conveyor belt $u < u_c$ and becomes proportional to $(R-kd)^{3/2}$ when $u > u_c$. Further experiments of granular flow on conveyor belt were performed by
Aguirrie \emph{et al.} \cite{Aguirre10,Aguirre12}.  Using the DEM \cite{Cundall1979}, microdynamic variable distributions of the granular flow in 3D cylindrical hoppers have been investigated by Zhu \emph{et al.} \cite{Zhu2004}. Under general gravity $g^*$, Dorbolo \emph{et al.} measured the mass flow rate which depends on the square root of the gravity $g^*$\cite{Dorbolo2013}.

 \begin{figure}[b]
\includegraphics[width=3 cm,height=4 cm]{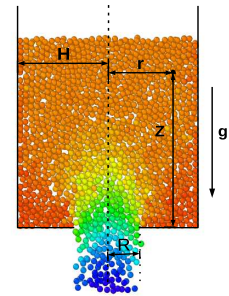}% Here is how to import EPS art
\caption{\label{fig:schematic} Schematic diagram of granular flow through an aperture.}
\end{figure} 
 
In this Letter, we provide a mechanism to understand the dependences of flow rate $W$ on the radius $R$ of apertures in hoppers. A schematic diagram of hopper is shown in Fig.\ref{fig:schematic}. At any position of a hopper, an average velocity ${\bm v}$ of granular flow can be defined. Because of the symmetry of hopper, vertical velocity $v_{\rm z}$ depends on $r$ and $z$. Furthermore,  $v_{\rm z}$ should be related to $R$ and the radius of hopper $H$. For hoppers with $H\gg R$, the dependency  of $v_z$ on $H$ can be neglected and we have $v_{\rm z} = v_{\rm z} (r,z;R)$. 

Using DEM and GPU program, granular flows in hopper have been simulated  \cite{Lin2015,Tian2015}. The correlation of velocities in different positions of hopper can be calculated from the simulation data. It was found that the velocities in the region above the aperture are correlated strongly. For a finite system near its critical point, the system is correlated strongly and there is 
the finite-size scaling \cite{Fisher1972Scaling}. Similar to the finite-size scaling of critical phenomena, we propose a size scaling form of the vertical  velocity field as
\begin{eqnarray}
\label{vscaling}
v_{\rm z} (r,z;R)=v_{\rm z} (0,0;R)f\left( r/R_{\rm r},z/R_{\rm z}\right),
\end{eqnarray}
where $R_{\rm r}$ and $R_{\rm z}$ are the radial and vertical sizes of the region. Because of the granule diameter $d$, we anticipate that $R_{\rm r}= R- 0.5 d$\cite{Zhang1991} . The scale length $R_{\rm z}= R +k_2d$, where $k_2$ is the only parameter to be determined.

During a time period $\delta t$, granules at $(r,z)$ move in average to $(r,z+v_z(r,z;R)\delta t)$ and have a velocity change $\delta v_z = v_z (r,z+v_z(r,z;R)\delta t;R)-v_z(r,z;R)$. We can calculate an effective acceleration as
\begin{equation}
a_{\rm eff}(r,z;R)\equiv \lim_{\delta t \to 0}\frac{\delta v_z}{\delta t }= \frac{\partial v_z(r,z;R)}{\partial z} v_z(r,z;R)\;.
\end{equation}
According to Eq.\ref{vscaling}, we get 
\begin{equation}
a_{\rm eff}(r,z;R)= \frac {v^2_z(0,0;R)}{R_{\rm z}} \theta \left( r/R_{\rm r},z/R_{\rm z}\right)\;,
\end{equation}
where 
\begin{eqnarray}
\theta \left(\bar r, \bar z \right)&=& f (\bar r, \bar z) f_z^{(1)} (\bar r,\bar z) \;,\\
f_z^{(1)} (\bar r,\bar z) &=& \frac {\partial f (\bar r,\bar z)}{\partial \bar z}\;.
\end{eqnarray}
At the center of aperture, the scaling function $f(0,0)=1$ and the effective acceleration
\begin{equation}
\label{aeff0}
a_{\rm eff} (0,0;R)= \frac {v^2_z(0,0;R)}{R_{\rm z}} f_z^{(1)} (0,0).
\end{equation}

When hopper radius $R\gg d$, granules can fall freely at aperture center and there is the boundary condition
\begin{equation} 
\label{bou}
a_{ eff}(0,0;R)=-g.
\end{equation}
We can then get the velocity
\begin{equation}
\label{eq:3} 
v_z(0,0;R)=\sqrt{g\lambda} \; R_z^{1/2},
\end{equation}
where $\lambda = -1/f_z^{(1)} (0,0)$.

For a $D$-dimnsional hopper, mass flow rate $W$ can be calculated as
\begin{equation}
W =\int_0^{R_{\rm r}}v_z(r,0;R) \rho S_{D-1} r^{D-2}  d r \;,
\end{equation}
where $S_1=2$ and $S_2 = 2\pi$. Using Eqs.(\ref{vscaling}) and (\ref{eq:3}), we obtain
\begin{equation}
\label{wscaling}
W = C_1\rho \sqrt{g }R_r^{D-1} R_z^{1/2}\;,
\end{equation}
with $C_1 = S_{D-1} [-f_z^{(1)} (0,0)]^{-1/2} \int_0^1 f(x,0)x^{D-2}  d x $.
In asymptotical case $R \gg d$, our Eq.(\ref{wscaling}) becomes  $W \simeq C_1\rho\sqrt{g}R^{D-\frac 1 2}$ and is in agreement with the Beverloo law.

To test the size scaling of velocity field above, we have made a series of simulations about granular flow throgh an aperture with the model of Ref.\cite{Lin2015}.  The granule has the diameter $d=1$ mm and density $\rho=2500$ kg/m$^3$.  Further parameters are listed in Table 1. 

\begin{table}[h]
\caption{\label{tab:table1}% 
Parameters of granular model.}
\centering
\begin{ruledtabular}
\begin{tabular}{lcdr}
\textrm{Quantity}&
\textrm{Symbol}&
\textrm{Value}\\
\colrule
Poisson's ratio &  $\nu$ & 0.2\\
Friction of spheres &  $\mu$ & 0.5\\
Coefficient of restitution & $\varepsilon$ & 0.8 \\
Shear modulus (Pa) & $G$ & 3.0\times 10^{10}\\
Elastic modulus  (Pa) & $E$ & 7.2\times 10^{10}\\
Density of spheres (kg/m) $^3$ & $\rho$ & 2.5 \times 10^{3} \\
Diameter of spheres (m) & $ d $ &  1.0 \times 10^{-3} \\
\end{tabular}
\end{ruledtabular}
\end{table}

To check the size scaling in general dimensionality,  granular flows both in 3D and 2D hoppers have been simulated. In 3D hoppers with radius $H=60d$, $N=4400000$ granules are studied for apertures with radius $R=5d, 10d, 15d, 20d, 25d$. The snapshots of positions and velocities of granules are taken every 100000 steps with step $\tau =5.0 \times 10^{-7}$ seconds. In 2D hoppers with the radius $H=100d$,  $N=200000$ granules have been simulated through apertures with radius $R=8d, 10d, 15d, 20d, 25d$.  The snapshots of 2D hoppers are taken every 20000 steps.

The vertical velocity field $v_z (r,h;R)$ can be obtained by two averages of simulation data. From a snapshot, vertical velocities of the granules inside a region with radius from $r-0.5 d$ to $r+0.5 d$ and height from $h-0.5 \delta_h$ to $h+0.5 \delta_h$ are known and their averages can be calculated. After the second average over snapshot, vertical velocity field $v_z (r,h;R)$ can be obtained. We chose $\delta_h=d, 1.5 d$ for 3D and 2D hopper, respectively.

\begin{figure}[hb]
\includegraphics[width=8cm,height=6cm]{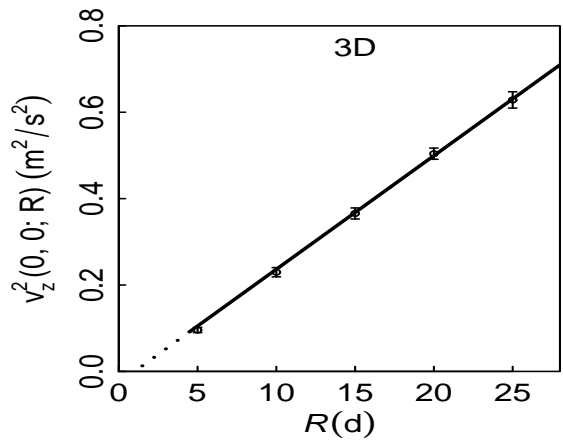}% Here is how to import EPS art
\caption{\label{fig:v0} $v^2_z(0,0;R)$ with respect to $R$ in 3D hoppers.  The solid line has the slope $g \lambda =26.3 m/s^2$ and its intersection point with $R$-axis gives $k_2 = -1.0$.}
\end{figure}

The simulation data of $v^2_z (0,0;R)$ for 3D hoppers are shown with respect to $R$ in Fig.\ref{fig:v0}. Good agreement of simulation data with the size scaling form $v^2_z (0,0;R)=g\lambda  (R+k_2 d)$ are found with $g\lambda =26.3 m/s^2 $ and $k_2=-1.0$. 

\begin{figure}
\includegraphics[width=8cm,height=11cm]{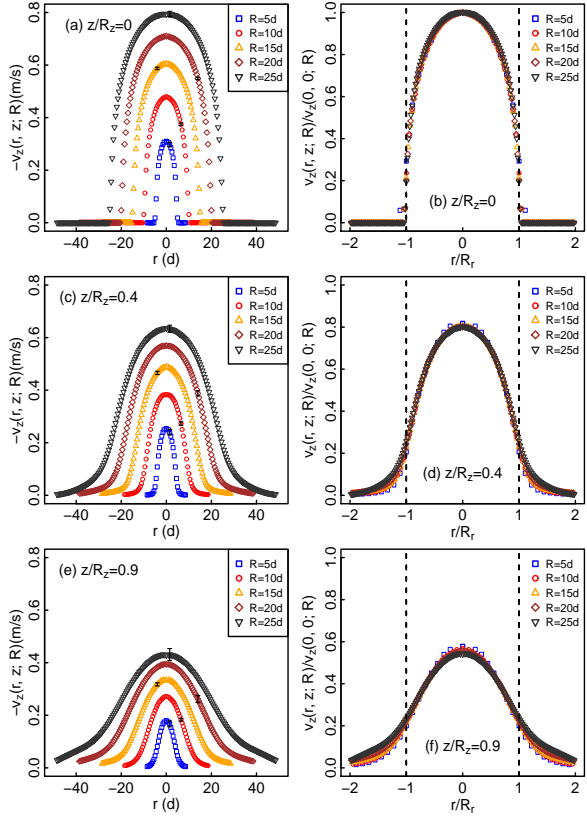}% Here is how to import EPS art
\caption{\label{fig:rvz3d} Vertical velocity field $v_z (r,z;R)$ in 3D hopper with respect to radial distance $r$ at different $z$ and $R$ with: (1) $z/R_{\rm z}=0$ in Fig.(a) ; (2) $z/R_{\rm z}=0.4$ in Fig.(c); (3) $z/R_{\rm z}=0.9$ in Fig.(e). The scaled vertical velocity field $v_z (r,z;R)/v_z (0,0;R)$ with respect to $r/R_{\rm r}$ at: (1) $z/R_{\rm z}=0$ in Fig.(b); (2) $z/R_{\rm z}=0.4$ in Fig.(d); (3) $z/R_{\rm z}=0.9$ in Fig.(f).}
\end{figure}

\begin{figure}
\includegraphics[width=8cm,height=11cm]{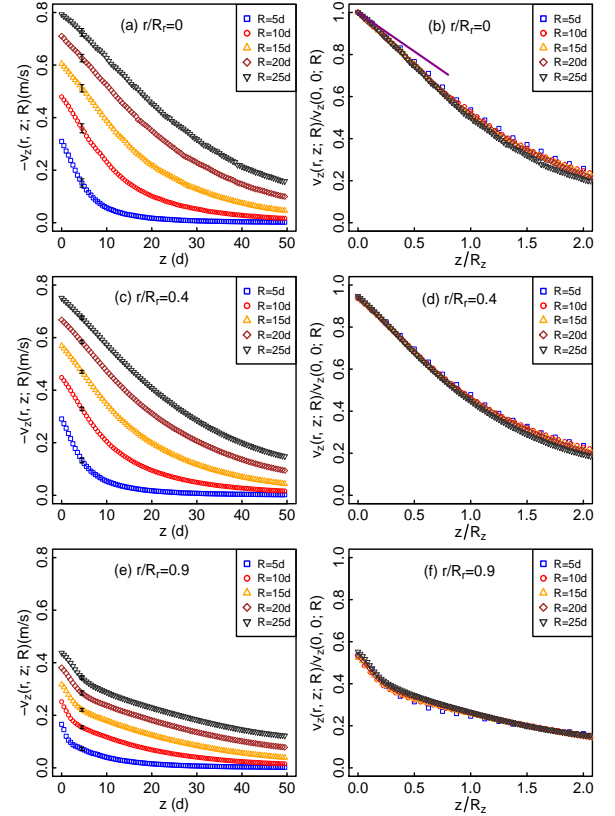}% Here is how to import EPS art
\caption{\label{fig:hvz3d}Vertical  velocity field $v_z (r,z;R)$ in 3D hopper with respect to height $z$ at different $r$ and $R$ with: (1) $r/R_{\rm r}=0$ in Fig.(a); (2) $r/R_{\rm r}=0.4$ in Fig.(c); (3) $r/R_{\rm eff}=0.9$ in Fig.(e). The scaled vertical velocity field $v_z (r,z;R)/v_z (0,0;R)$ with respect to $z/R_{\rm z}$ at: (1) $r/R_{\rm eff}=0$ in Fig.(b); (2) $r/R_{\rm eff}=0.4$ in Fig.(d); (3) $r/R_{\rm eff}=0.9$ in Fig.(f). The straight line in Fig.(a) has the slope in agreemet with that of Fig.(\ref{fig:v0}).}
\end{figure}

On the left side of Fig.\ref{fig:rvz3d},  simulation data of vertical velocity fields $v_z (r,z;R)$ for different $z$ and $R$ are plotted with respect to $r$.  
After rescaling $v_z (r,z;R)$ by $v_z (0,0;R)$ and $r$ by $R_r = R-0.5 d$, the different curves of the same scaled height $z/R_{\rm z}$ with $R_{\rm z}=R-1.0 d$ collapse together and are shown on the right side Fig.\ref{fig:rvz3d}. 

On the left side of Fig.\ref{fig:hvz3d}, simulation data of $v_z (r,z;R)$ are shown with respect to $z$. As a function of the scaled height $z/R_{\rm z}$, the scaled velocity field $v_z (r,z;R)/v_z (0,0;R)$ is shown on the right side of Fig.\ref{fig:hvz3d}. The diffferent curves on the left side collapse together.

As shown in Fig.\ref{fig:hvz3d}(b), the solid line is drawn with the slope  $f_z^{(1)} (0,0)=-\lambda^{-1}$ with $\lambda$ determined in Fig.\ref{fig:v0}. Good agreement is found and the boundary condition $a_{\rm eff}(0,0;R)= -g$ is confirmed by our simulation data.

From Figs.\ref{fig:rvz3d} and \ref{fig:hvz3d}, we can conclude that the size scaling form in Eq.(\ref{vscaling}) is confirmed by our simulation data for 3D hopper.

\begin{figure}[hb]
\includegraphics[width=8cm,height=6cm]{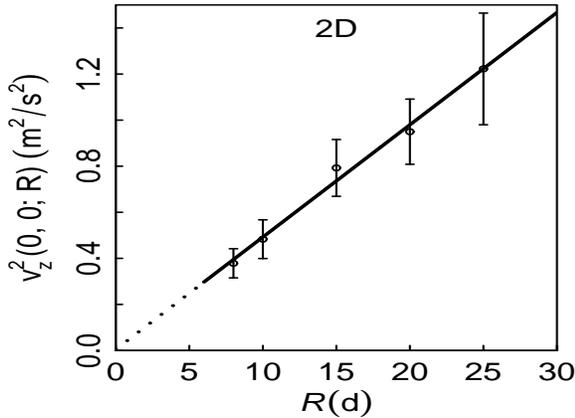}% Here is how to import EPS art
\caption{\label{fig:v02} $v^2_z(0,0;R)$ of 2D hoppers with respect to $R$. The solid line has the slope $g \lambda =48.7 m/s^2$ and its intersection point with $R$-axis gives $k_2 = 0.1$.
}
\end{figure}

For 2D hopper, $v_z^2 (0,0;R)$  is plotted with respect to $R$ in Fig.\ref{fig:v02}.  We find also a linear $R$-dependence of $v^2_z (0,0;R)$ with $\lambda g = 48.7 m/s^2$. From the intersection of the solid line with the $R$-axis in Fig.\ref{fig:v02},  we get $k_2 = 0.1$.

\begin{figure}
\includegraphics[width=8cm,height=11cm]{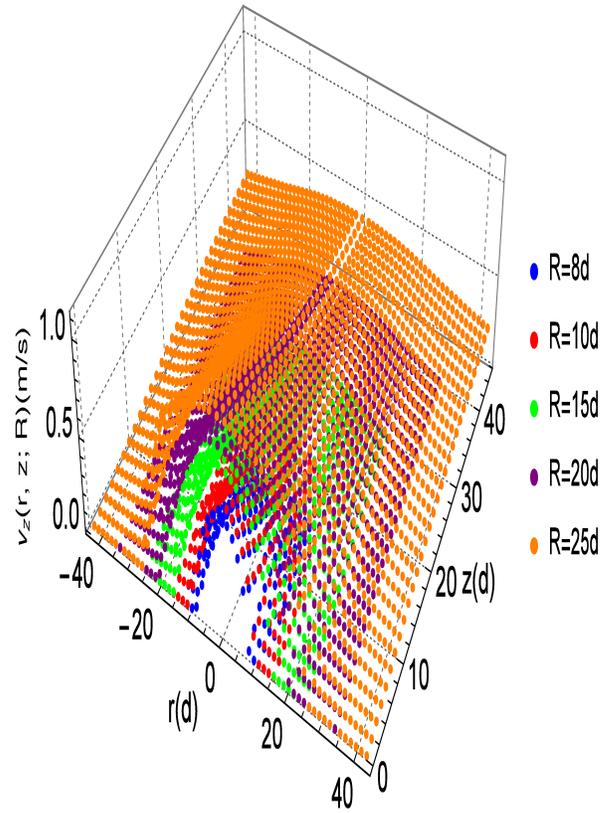}% Here is how to import EPS art
\caption{\label{fig:vrz2d} Vertical velocity fields $v_{\rm z} (r,z;R)$ of 2D hoppers with different aperture radius are plotted with respect to $r$ and $z$.}
\end{figure}

\begin{figure}
\includegraphics[width=8cm,height=11cm]{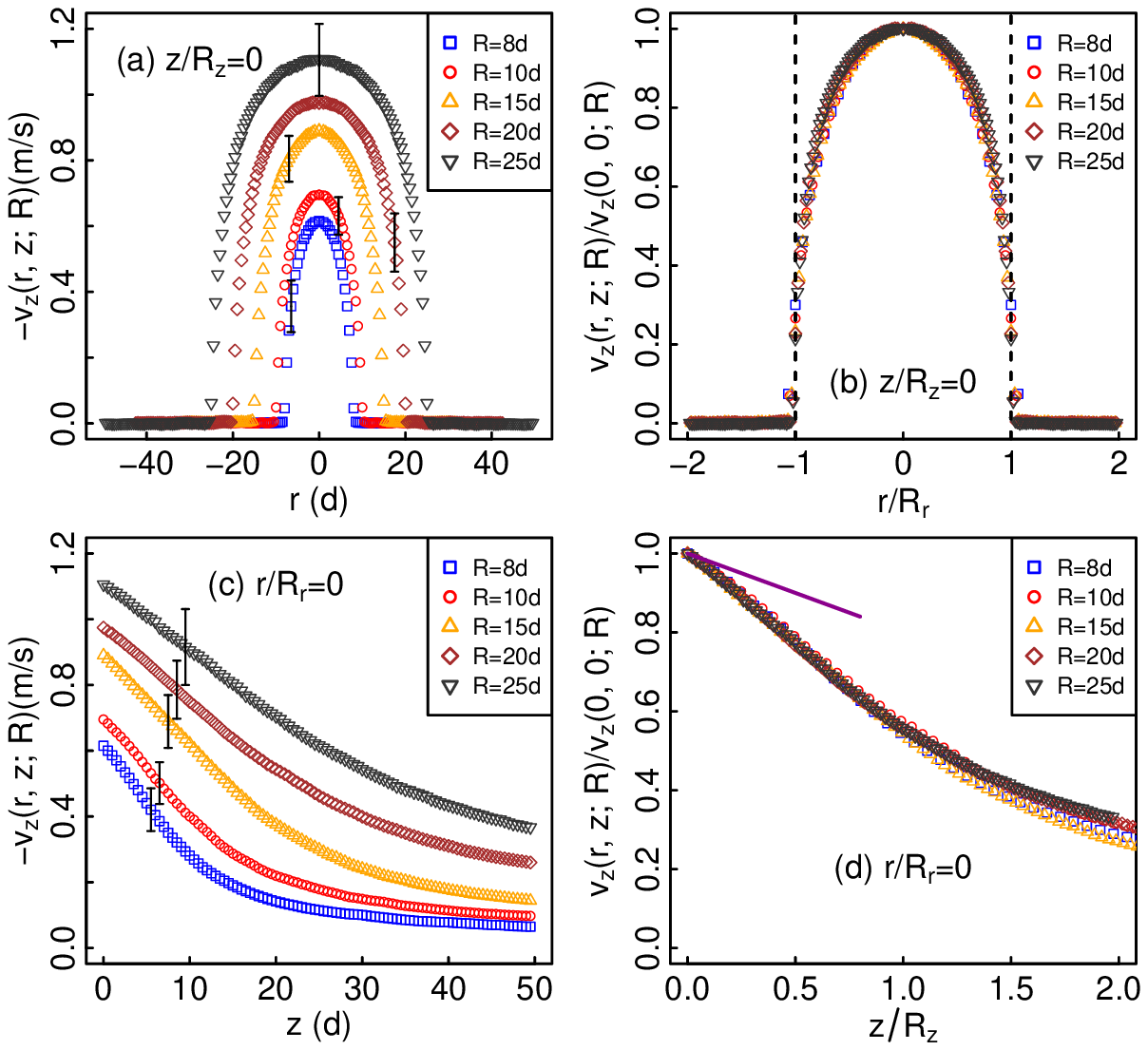}% Here is how to import EPS art
\caption{\label{fig:rvz2d} Vertical  velocity field $v_z (r,z;R)$ of 2D hopper is plotted  versus $r$ at $z=0$ in (a) and versus $z$ at $r=0$ in (c). The scaled velocity field $v_z (r,z;R)/v_z (0,0;R)$ is presented with respect to $r/R_{\rm r}$ in (b) and $z/R_{\rm z} $ in (d).}
\end{figure}

Vertical velocity fields $v_{\rm z} (r,z;R)$ of 2D hoppers with aperture radius $R=8d, 10d, 15d, 20d, 25d$ are shown in Fig.\ref{fig:vrz2d}.  We can see the obvious $R$-dependence of $v_z (r,z;R)$. According to Eq.\ref{vscaling}, the scaled velocity field $v_{\rm z} (r,z;R)/v_{\rm z} (0,0;R)$ is plotted in Fig.\ref{fig:rvz2d} as a function of  $r/R_{\rm r}$ and $z/R_{\rm z}$ with $R_{\rm r}=R-0.5d$ and $R_{\rm z}= R+0.1d$.  The different curves of $v_{\rm z} (r,z;R)$ in Fig.\ref{fig:vrz2d} collapse together in Fig.\ref{fig:rvz2d}. In Fig.\ref{fig:rvz2d}, these results are presented especially at $z=0$ and $r=0$. The size scaling form of Eq.\ref{vscaling} is also confirmed by our simulation data of 2D hoppers.

\begin{figure}
\includegraphics[width=8cm,height=11cm]{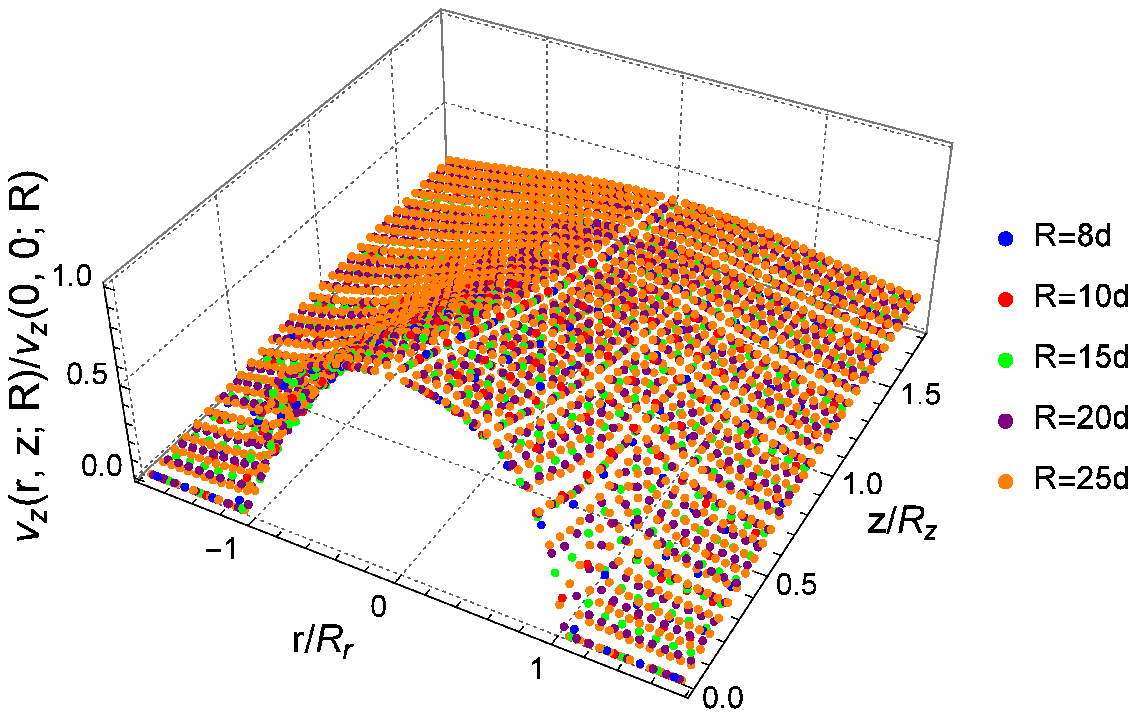}% Here is how to import EPS art
\caption{\label{fig:svrz2d} Scaled vertical velocity field $v_z (r,z;R)/v_z (0,0;R)$ of 2D hoppers is plotted as a function of $r/R_{\rm r}$ and $z/R_{\rm z}$ with $R_{\rm r}=R-0.5d$ and $R_{\rm z}= R+0.1d$.}
\end{figure}

The normalized velocity $v_z (r,0;R)/v_z (0,0;R)$ of 2D silo has been investigated experimentally by A. Janda \emph{et al.} \citep{Janda}.  As a function of $r/R$, their experimental data of different $R$  collapse together. For both 2D and 3D silos, S.M. Rubio-Largo \emph{et al.}\citep{Rubio} studied experimentally the normalized effective acceleration $a_{\rm eff} (0,z;R)/g$ as a function of $z/R$ and there is no sign of existence of the free-fall arch. 

Under general gravity $g^*$, the boundary condition becomes  $a_{\rm eff}(0,0;R)= - g^*$ and $\sqrt{g}$ of the mass flow rate in Eq.\ref{wscaling} is replaced by $\sqrt{g^*}$, which is in agreement with experimental results of Dorbolo \emph{et al.} \cite{Dorbolo2013}. 

For granular flows on conveyor belt, we have the boundary condition $v (0,0;R)=u$ when $u$ is small. Then the mass flow rate can be calculated as
$W = C_2 \rho u R_{\rm r}^{D-1}$ with $C_2 = S_{D-1}  \cdot\int_0^1 f(x,0)x^{D-2}  d x$. This result  is in agreement with the experimental results \citep{Bao2003,Aguirre10,Aguirre12}. If the velocity of conveyor belt is larger than a critical value, there is no boundary condition for $v (0,0;R)$.  The boundary condition now becomes $a_{\rm eff} (0,0;R)= - a_f$ with $a_f$ related to the friction force between granules and the belt and $W \propto \rho \sqrt{a_f} R_{\rm r}^{D-1} R_{\rm z}^{1/2}$. Therefore, the $R$-dependence of flow rate $W$ on conveyor belt with $D=2$ switches from $R_{\rm r}$ to $R_{\rm r} R_{\rm z}^{1/2}$ with the increase of belt velocity, as found in the experiment \cite{Bao2003}.

In summary, a size scaling form of vertical velocity field in granular flow through an aperture with radius $R$ is proposed as $v_{\rm z} (r,z;R)=v_{\rm z} (0,0;R)f(r/R_{\rm r},z/R_{\rm z})$ in the region above the aperture. The length scales $R_{\rm r}=R-0.5 d$ and $R_{\rm z}=R + k_2 d$, where $k_2$ is a parameter to be determined. From $v_z (r,z;R)$, we can get an effective acceleration $a_{\rm eff} (r,z;R)$ and its size scaling form $a_{\rm eff} (r,z;R)= v_{\rm z}^2(0,0;R)R_{\rm z}^{-1}  f (r/R_{\rm r}, z/R_{\rm z}) f_z^{(1)} (r/R_{\rm r}, z/R_{\rm z})$, where $f_z^{(1)}$ is the partial derivative of scaling function . For granular flow under gravity $g$, there is a boundary condition $a_{\rm eff} (0,0;R)=-g$ which gives rise to $v_{\rm z} (0,0;R)=\sqrt{\lambda g R_{\rm z}}$ with $\lambda = -1/f_z^{(1)} (0,0)$. Then we get the flow rate $W = C_1 \rho \sqrt{g} R_{\rm r}^{D-1} R_{\rm z}^{1/2}$, which agrees with the Beverloo law when $R \gg d$. For granular flow on conveyor belt with small speed $u$, there is a boundary condition $v_z (0,0;R)=u$ and the flow rate $W = C_2 \rho u R_{\rm r}^{D-1}$. When $u$ is larger than a critical value, there will be a fixed friction force acting on granules by conveyor belt and the boundary condition now is related the effective acceleration.
This can explain the switch of the $R$-dependence of flow rate $W$ with conveyor belt speed $u$ as observed in the experiment \cite{Bao2003}.
  
Using DEM and GPU program, granular flows under gravity have been simulated for 3D and 2D hoppers and different radius of the aperture. Our simulation data confirm the size scaling form of $v_{\rm z} (r,z;R)$ proposed above. Furthermore, the $R$-dependence of the vertical velocity  at the center of aperture $v_{\rm z} (0,0;R)=\sqrt{\lambda g R_{\rm z}}$ is in agreement with our simulation data.

This work is supported by the National Magnetic Confinement Fusion Science Program of China under Grant No. 2014GB104002,
the Strategic Priority Research Program of the Chinese Academy of Sciences under Grant No. XDA03030100, and the National natural Science Foundation of China under Grant No. 11421063.

%\bibliography{reference}% Produces the bibliography via BibTeX.

%

\end{document}